\documentclass[a4paper,10pt]{article}
\usepackage{graphicx}
\usepackage{amsmath,amsfonts,amsthm,amssymb}

\swapnumbers %Theorems, lemmas, etc. numbered at the left.

\newtheorem{lemma}{Lemma}[section]

\newtheorem{theorem}[lemma]{Theorem}
\newtheorem{proposition}[lemma]{Proposition}
\newtheorem{example}[lemma]{Example}

\theoremstyle{definition} %Set Following proclamations in normal text.
\newtheorem{definition}[lemma]{Definition}
\newtheorem{remark}[lemma]{Remark}

\begin{document}

\title{Granger causality on horizontal sum of Boolean algebras}

% \thanks{The authors kindly announce the support from the Science and Technology Assistance Agency under contract No. APVV-14-0013, and from the VEGA grant agency, grant numbers 1/0710/15 and 2/0069/16.}

\author{
	M\'aria Bohdalov\'a
	\footnote{Dept. of Inf. Systems, Faculty of 		Management 		Comenius University, Bratislava, Slovakia, maria.bohdalova@fm.uniba.sk}
	 \and
     Martin Kalina
     \footnote{Dept. of Mathematics, Faculty of Civil Engineering, Slovak University of Technology,     Bratislava, Slovakia,   kalina@math.sk}
     \and 
     O\v{l}ga N\'an\'asiov\'a
     \footnote{Institute of Computer Science and Mathematics, Faculty of Electrical Engineering and Information Technology, Slovak University of Technology,  Bratislava, Slovakia, olga.nanasiova@stuba.sk}
 }
\maketitle
\date{}
%\end{document}

\begin{abstract}
The intention of this paper is to discuss the mathematical model of causality introduced by C.W.J. Granger in 1969. The Granger's model of causality has become well-known and often used in various econometric models describing causal systems, e.g., between commodity prices and exchange rates. 

Our paper presents a new mathematical model of causality between two measured objects. We have slightly modified the well-known Kolmogorovian probability model. In particular, we use the horizontal sum of  set $\sigma$-algebras instead of their direct product.
\end{abstract}

\section{Introduction}
\label{intro}
In the past there were several attempts to model  causality. Yet, there exists no  mathematical definition of this notion. Loosely understood is such a relationship between a cause $\xi$ and its effect $\eta$. The usual dependence relationship between $\xi$ and $\eta$ is symmetric (often expressed by a type of correlation). However, there are situations where we have non-symmetric dependence between variables, e.g., the past of a time-series may influence its future, but not vice-versa. Or in hydrology, high water level in a river may cause high water level dawn the river flow, but usually not vice-versa. Thus, we need also a non-symmetric relationship between a cause and its effect. 

The aim of this paper is to provide a mathematically rigorous model of non-symmetric causality. It will be based on Granger's approach to causality \cite{G}. From the algebraic point of view, our model will be based on horizontal sums of Boolean algebras. 
A preliminary version of our model was published in \cite{BKN1}. This is an extended version of the paper \cite{BKN1}.

Our paper is organized as follows: in section \ref{sec:1} we provide a historical overview of various models of causality. In section \ref{sec:2} we recall the Granger's definition of causality for stationary time-series. Section \ref{sec:3} contains basic definitions and some known facts on orthomodular lattices. In section \ref{sec:4} we provide a theoretical background for our model of causality. It has two subsections:\\
-- \ref{sec:4-1} contains a theory of bivariate states on orthomodular lattices,\\
-- \ref{sec:4-2} contains an approach that enables to add non-compatible observables.

Finally, section \ref{sec:5} contains our model of causality on horizontal sums of Boolean algebras. It has again two subsections:\\
-- \ref{sec:5-1} provides a comparison of random vectors on Boolean algebras and of vectors of observables on horizontal sums of Boolean algebras,\\
-- \ref{sec:5-2} is devoted to the Granger's model of causality modified for horizontal sums of Boolean algebras.

\section{Historical overview of modelling causality in economics}
\label{sec:1}
In the 20th century  causal inference was frequently associated with multiple correlation and regression. As it is well known, the regression of Y on X produces coefficient estimates that are not the algebraic inverses of those  produced from the regression of X on Y. Although regressions may have a natural causal direction, there is nothing in the data on their own that reveals which direction is the correct one – each of them is an equally appropriate rescaling of a symmetrical and non-causal correlation. This is a problem of \emph{observational equivalence}. For example, we can mention the problem of econometric identification: \emph{how to distinguish a supply curve from a demand curve}. A standard solution to this identification problem is to look for additional causal determinants that discriminate between otherwise simultaneous relationships. Possible solution of this problem gives us the language of exogenous  and endogenous variables. Exogenous variables can also be regarded as the causes of the endogenous ones \cite{Hoo08}.

 In the 1930s Jan Tinbergen \cite{Tin} introduced structural models in modern econometrics. These models  express causality in a diagram that uses arrows to indicate causal connections among time-dated variables. Another approach is known as process analysis. Process analysis emphasizes the asymmetry of causality, typically grounded  in Hume's criterion of temporal precedence \cite{Mor}. Wold's process analysis belongs to the time-series tradition that ultimately produces Granger causality and vector autoregression. The Wold's approach relates causality to the invariance properties of the structural econometric model. This approach emphasizes the distinction between endogenous and exogenous variables and the identification and estimation of structural parameters. Herbert Simon \cite{Sim} has shown that causality could be defined in a structural econometric model, not only between exogenous and endogenous variables, but also among the endogenous variables themselves. And he has shown that the conditions for a well-defined causal order are equivalent to the well-known conditions for identification \cite{Hoo08}.

Hans Reichenbach \cite{GE12,Rei56}, taken the idea that simultaneous correlated events must have prior common causes, tried to use them to infer the existence of unobserved and unobservable events and to infer causal relations from statistical relations. Reichenbach's common cause principle is a time-asymmetric principle that can be formulated as follows: simultaneous correlated events have a prior common cause that screens off the correlation. It means, if simultaneous values of quantities $A$ and $B$ are correlated, then there are common causes $C_1,C_2,\ldots,C_n$ such that conditioned upon any combination of values of these quantities at an earlier time, the values of $A$ and $B$ are probabilistically independent, see \cite{Af10,Uff99}. Reichenbach's common cause principle was adopted by Penrose and Percival \cite{pp62} into the law of conditional independence and by Spirtes et al. \cite{SGS93} into causal Markov condition. 

Some open problems concerning Reichenbachian common cause systems are formulated and solved by Hofer-Szabó and Rédei in many papers. Hofer-Szabó and Rédei \cite{HR06} have shown that given any non-strict correlation in $(\Omega ,\mathcal S, P)$ and given any finite natural number $n>2$, the probability space $(\Omega ,\mathcal S, P)$ can be embedded into a larger probability space in such manner that the larger space contains a Reichenbachian common cause system of size
$n$ for the correlation.

Another approach, given by Clive W. J. Granger \cite{G},  introduces the data-based concept without direct reference to background economic theory. This concept has become a fundamental notion for studying dynamic relationships among time series. Granger's causality is an example of the modern probabilistic approach to causality, and it is a natural successor to Hume (see, e.g., \cite{Sup70}). Where Hume requires constant conjunction of cause and effect, probabilistic approaches are content to identify cause with a factor that raises the probability of the effect: $A$ causes $B$ if $P(B|A) > P(B)$, where the vertical ‘$|$’ indicates ‘conditioned on’. The asymmetry of causality is secured by requiring the cause $(A)$ to occur before the effect $(B)$ (see \cite{Hoo08}). But the probability criterion is not enough on its own to produce asymmetry since $P(B|A) > P(B)$ implies $P(A|B) > P(A)$.

Granger's causality helps us to understand and measure the relative roles of different causal systems, e.g., between commodity prices and exchange rates. Granger causality has important implications in financial decision making, especially for market participants with short horizons. From a macroeconomic perspective, this can also be useful for interpreting exchange rate movements, financial market monitoring and monetary policy. Basic economic reasoning on currency demand suggests that the currencies of countries whose exports depend heavily on a particular commodity should be strongly influenced by its price, so commodity price movements should lead (Granger-cause) exchange rate movements (macroeconomic/trade mechanism).

\section{Introduction to the Granger's model of causality}
\label{sec:2}
In statistics, the notion of causality is usually identified with a kind of stochastic dependence. Of course, this dependence (e.g. between two random variables) is a symmetric notion. In \cite{G}, Granger defined a causality between two stationary time-series ${\mathbf X}=\{X_t\}_{t\in{\mathbb Z}}$ and ${\mathbf Y}=\{Y_t\}_{t\in{\mathbb Z}}$ in a non-symmetric way. There are two basic principles upon which this notion of causality (and a relationship between a cause and its effect) is based.
\begin{itemize}
	\item Cause always happens prior to its effect.
	\item Cause makes unique changes in the effect. In other words, the causal series contains
unique information about the effect series that is not available otherwise.
\end{itemize}
The precise definition of Granger's causality is the following:
\begin{definition}[\cite{G}]\label{def-G}
Denote by $U_{t-1}$ all information of the universe that is accumulated by time $t-1$ and $\bar{X}_{t-1}=\{X_{t-i}\}_{i=1}^{\infty}$ the past of ${\mathbf X}$ by time $t-1$. 
Let $\tilde{U}_{\bar{X}_{t-1}}$ denote all the information of the universe that is accumulated by time $t-1$ apart from the past $\bar{X}_{t-1}$. Then if $\sigma^2(Y_t|U_{t-1})<\sigma^2(Y_t|\tilde{U}_{\bar{X}_{t-1}})$, we say that ${\mathbf X}$ is causing ${\mathbf Y}$. %, denoted by ${\mathbf X}\stackrel{c}{\rightarrow}{\mathbf Y}$.
\end{definition}

\begin{remark}\rm\quad
\begin{enumerate}
\item Since we assume stationarity of the time series ${\mathbf X}$ and ${\mathbf Y}$, the condition $\sigma^2(Y_t|U_{t-1})<\sigma^2(Y_t|\tilde{U}_{\bar{X}_{t-1}})$ is fulfilled for all $t$ whenever it is fulfilled for one $t$. This means that the past of the time series ${\mathbf X}$ influences the future of ${\mathbf Y}$ (expressed by means of the conditional variance in Definition \ref{def-G}).
\item As Granger pointed out in \cite{G}, we could skip the condition that the time series ${\mathbf X}$ and ${\mathbf Y}$ are stationary, but then the causality between ${\mathbf X}$ and ${\mathbf Y}$ would become time-dependent, i.e., it might occur that at some time stamps $t_i$ $X_{t_i}$ is causing $Y_{t_i}$ %we would have $X_{t_i}\stackrel{c}{\rightarrow} Y_{t_i}$ 
and at some other time stamps ${t_j}$, $X_{t_j}$ would not cause $Y_{t_j}$.
\end{enumerate}
\end{remark}

 In fact, this notion of causality is based on the Kolmogorovian conditional probability theory. Granger's theory is used especially in econometrics and finance to model one-sided dependencies, as we have mentioned in the historical overview. 

\section{Basic definitions and known facts on orthomodular lattices}
\label{sec:3}
We have already mentioned in Introduction that our model of causality works on horizontal sums of Boolean algebras. 
Since horizontal sums of Boolean algebras are special cases of orthomodular lattices, we recall some basic facts also on general orthomodular lattices.
 For more information on orthomodular lattices and their properties one can consult, e.g., \cite{D-P,Kalm-83,PtakPulm,Var}.
\begin{definition}\label{def:OML}
Let $(L,{\mathbf 0}_L,{\mathbf 1}_L,\vee ,\wedge ,\,' )$ be a
lattice with the greatest element ${\mathbf 1}_L$ and the
least element ${\mathbf 0}_L$.  Let
 $':L\to L$  be a unary operation on $L$ with the following properties:
\begin{enumerate}
\item for all $a\in L$ there is a unique $a'\in L$ such that
$(a')' =a$ and $a\vee a'= {\mathbf 1}_L$;%
\item if $a,b\in L$ and $a\le b$ then $b'\le a' $; %
\item if $ a,b\in L$ and $a\le b$ then $b=a\vee (a'\wedge b)$
(orthomodular law).
\end{enumerate}
Then $(L,{\mathbf 0}_L,{\mathbf 1}_L,\vee ,\wedge ,\,' )$ is
said to be \emph{an orthomodular lattice}.
\end{definition}
In this paper, for the sake of brevity, we will write briefly an
orthomodular lattice $L$, skipping the operations whenever it will
not cause any confusion. In general, an orthomodular lattice is not distributive.
For arbitrary $a,b\in L$ just the following property is
guaranteed
\[
(a\wedge b)\vee (a\wedge b' )\leq a.
\]
On the other hand, if $L$ is distributive then it is a Boolean
algebra.
\begin{definition}\label{def:komort}
Let $L$ be an orthomodular lattice. Then elements $a,b\in
L$ are called
\begin{enumerate}
 \item[\rm (o1)]  \emph{orthogonal} ($a\perp b$) if $a\le b' $,
\item[\rm (o2)] \emph{compatible} ($a\leftrightarrow b $)
if $a=(a\wedge b)\vee (a\wedge b' )$ and $b=(a\wedge b)\vee (a' \wedge b )$.
\end{enumerate}
\end{definition}

 \emph{An orthomodular sub-lattice $L_1$
of $L$} is an orthomodular lattice such that $L_1\subset L$,
with operations inherited from $L$ and possessing the same
greatest and least elements ${\mathbf 1}_L$ and ${\mathbf
0}_L$, respectively. A distributive orthomodular
sub-lattice $\cal B$ is called a Boolean
sub-algebra of $L$.

Every orthomodular lattice $L$ is a collection of blocks \cite{R2}. A
\emph{block} is the maximal set of pairwise compatible elements of
$L$, i.e. $L=\bigcup_j B_j$, where blocks $B_j$ have operations
inherited from $L$. Each block in $L$ is a Boolean
algebra.
\begin{definition}[See, e.g., \cite{D-P}]\label{hor-sum}
Let $L$ be an orthomodular lattice with the greatest element ${\mathbf1}_L$ and the least element ${\mathbf0}_L$. Moreover, let $L=\bigcup\limits_{j\in J} B_j$, where $B_j$ are blocks in $L$ for all $j\in J$.  We say that $L$ is a horizontal sum of Boolean algebras $B_j$ if 
\[
B_i\cup B_j=\{{\mathbf0}_L,{\mathbf1}_L\}\quad\mbox{for all $i,j\in J$ such that $i\ne j$.}
\]
\end{definition}

\begin{example}\label{exam-OML}\rm
Let $L_1=\{{\mathbf0}_{L_1}, a,a',b,b',{\mathbf1}_{L_1}\}$, as sketched  on Fig.\ref{fig-1}. Then $L_1$ is an  orthomodular lattice whose blocks are ${B}_1=\{ {\mathbf0}_{L_1}, a,a',{\mathbf1}_{L_1}\}$ and ${B}_2=\{{\mathbf0}_{L_1},b,b',{\mathbf1}_{L_1}\}$. The lattice $L_1$ is the horizontal sum of Boolean algebras  ${B}_1$ and ${B}_2$,  since ${B}_1\cap{B}_2=\{{\mathbf0}_{L_1},{\mathbf1}_{L_1}\}$.

Let $L_2=\{{\mathbf0}_{L_2},a,b,c,d,e,a',b',c',d',e',{\mathbf1}_{L_2}\}$ such that $a'=b\vee c$, $b'=a\vee c$, $c'=a\vee b=d\vee e$, $d'=c\vee e$, $e'=c\vee d$. $L_2$ is sketched on Fig. \ref{fig-2}. The orthomodular lattice has also 2 blocks, namely ${B}_3=\{ {\mathbf0}_{L_2}, a,b,c,a',b',c',{\mathbf1}_{L_2}\}$ and ${B}_4=\{{\mathbf0}_{L_2},c,d,e,c',d',e',{\mathbf1}_{L_2}\}$. The lattice $L_2$ is not the horizontal sum of Boolean algebras  ${B}_3$ and ${B}_4$ since ${B}_3\cap{B}_4=\{{\mathbf0}_{L_2},c,c',{\mathbf1}_{L_2}\}\ne\{{\mathbf0}_{L_2},{\mathbf1}_{L_2}\}$. 

\begin{figure}
	\begin{minipage}[b]{0.45\textwidth}
		\includegraphics[height=4.3cm]{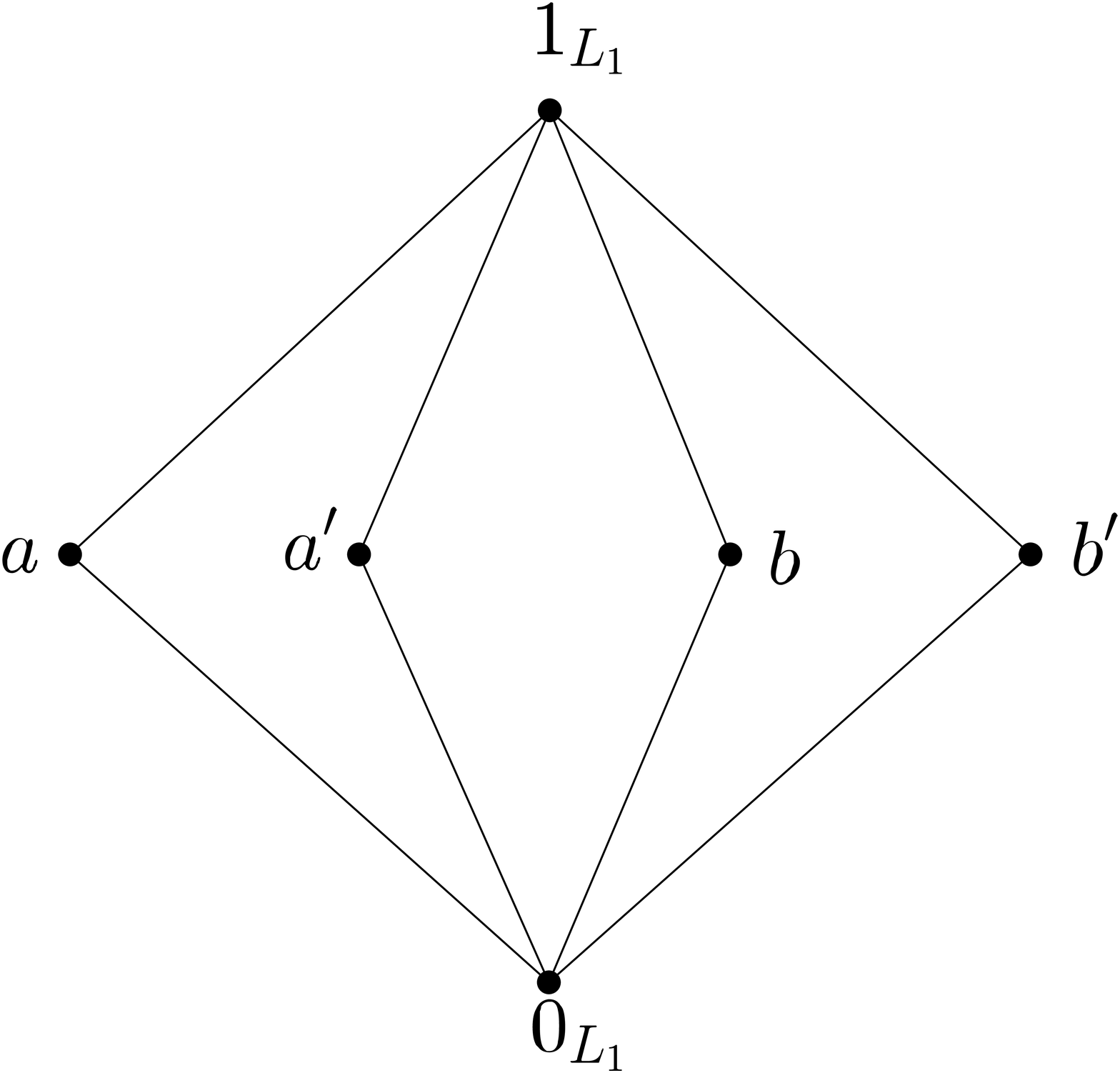}
		\caption{\label{fig-1}OML $L_1$}
	\end{minipage}
	\begin{minipage}[b]{0.45\textwidth}
		\includegraphics[height=4.3cm]{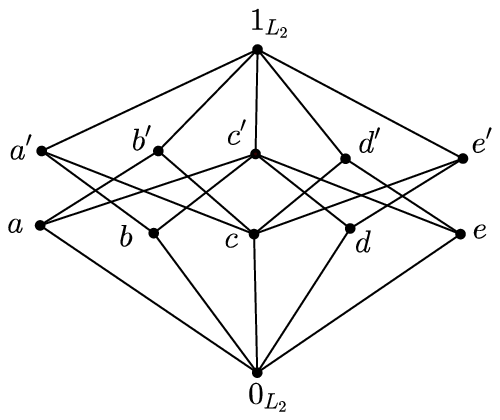}
		\caption{\label{fig-2}OML $L_2$}
	\end{minipage}
\end{figure}

\end{example}
\newpage
In this paper we will deal only with $\sigma$-complete orthomodular lattices $L$. Such  $\sigma$-complete orthomodular lattices are called orthomodular $\sigma$-lattices ($\sigma$-OML, for brevity).

\begin{definition}\label{def:state}
A map $m:L\to [0,1]$ is called \emph{a $\sigma$-additive state on
$L$}, if for arbitrary at most countable system of mutually
orthogonal elements $a_i\in L$, $i\in I\subset {\mathbb N}$, the
following holds
\[
m\left(\bigvee\limits_{i\in I}
a_i\right)=\sum\limits_{i\in I}m(a_i)
\]
and $m({\mathbf 1}_L)=1$.
\end{definition}

As it was proven by Greechie \cite{Gree}, there exist
orthomodular lattices with no state.

\begin{definition}\label{def:observ}
Let $L$ be a $\sigma$-OML. A $\sigma-$homomorphism $x$ from Borel
sets $\mathcal B({\mathbb R})$ to $L$, such that
 $x({\mathbb R})={\mathbf 1}_L$ is called \emph{an observable on $L$}.\\
By $\mathcal O$ we will denote  the set
of all observables on $L$.
\end{definition}

\begin{definition}\label{def:ranspec}
Let $L$ be a $\sigma$-OML and  $x$ be an observable on $L$. Then
\begin{enumerate}
\item[\rm (q1)] the set $R(x)=\{x(E); E\in\mathcal B({\mathbb R})\}$ is
called \emph{the range of the observable $x$ on $L$}; %
\item[\rm (q2)] the set $\sigma(x)=\cap\{E\in\mathcal B({\mathbb R});
x(E)={\mathbf 1}_L\}$ is called \emph{the spectrum of  an
observable}.
\end{enumerate}
\end{definition}

Directly from the properties of $\sigma $-homomorphism it follows
that $R(x)$ is a Boolean sub-$\sigma$-algebra of $L$
(e.g., \cite{PtakPulm,Var}).

\begin{definition}\label{def:comob}
Let $L$ be a $\sigma$-OML. Observables $x,y$ are called
\emph{compatible} ($x\leftrightarrow y $) if  $x(A)\leftrightarrow
y(B)$ for all $A,B\in\mathcal B({\mathbb R})$.
\end{definition}

\begin{theorem}[Loomis-Sikorski Theorem \cite{Var}]\label{L-S}
Let $L$ be a $\sigma$-OML and $x,y$ be compatible observables on
$L$. Then there exists a $\sigma$-homomorphism $h$ and real
functions $f,g$ such that $x(A) =h(f(A))$ and $y(A)
=h(g(A))$ for each $A\in\mathcal B({\mathbb R})$ (briefly $x
=h\circ f$ and $y =h\circ g$).
\end{theorem}

If $x\in\mathcal O$  and $m$ is a $\sigma$-additive state on $L$, then
$m_x(B)=m(x(B))$, $B\in\mathcal{B({\mathbb R})}$ is a probability
distribution of $x$.

Let $(\Omega ,\mathcal S,P)$ be a probability space. Then
$\mathcal S$ is a Boolean $\sigma$-algebra and $P$ is a $\sigma$-additive state.
Hence $\mathcal S$ is a $\sigma$-OML. Furthermore, if $\xi $ is a
random variable on $(\Omega ,\mathcal S, P)$, then $\xi^{-1}$ is
an observable. It means that, if we have an observable $x$ on a
$\sigma$-OML $L$, we are in the same situation as in the
classical probability space. We use only another language for the
standard situation. Problems  occur if we have more then just
one observable, and their ranges are not compatible.

\section{Causality and orthomodular lattices}
\label{sec:4}
In this section we show how it is possible to introduce causality on orthomodular lattices between observables. As a first step we need conditional states and joint distributions (s-maps).
 
\subsection{Bivariate states on orthomodular lattices}
\label{sec:4-1}
Conditional states and  s-maps were introduced in
\cite{Ns,Nc} resp.,
 and their properties were studied for example in \cite{NP08}. For a given $\sigma$-OML $L$ with a $\sigma$-additive state,  $L_{0}$ will  denote the set of all elements $a\in L$ for which there exists a $\sigma$-additive state $m_a$ such that $m_a(a)=1$. In this paper we will assume that
\begin{equation}\label{L-0}
 L_0=L\setminus\{{\mathbf 0}_L\}.
\end{equation}

\begin{definition}\label{def:constate}
Let $L$  be a $\sigma$-OML. Let $f:L\times L_0\to [0,1]$ be a
function fulfilling the following
\begin{enumerate}
\item[\rm(c1)] for each $a\in L_0$  $f(.|a)$ is a $\sigma$-additive state on $L$;
\item[\rm(c2)] for each $f(a|a)=1$;%
\item[\rm(c3)] for mutually orthogonal (at most countably many)
elements $a_1,a_2,...\in L_0$ and for all $b\in L$ the following is satisfied
$$f\left(b\left|\bigvee_i \right.a_i\right)=\sum_if(b|a_i)
f\left(a_i\left|\bigvee_ia_i\right.\right).$$
\end{enumerate}
Then $f$ is called \emph{a conditional state on $L$}.
\end{definition}

\begin{remark}\rm
Assume that $L$ is an orthomodular lattice fulfilling \eqref{L-0}. If we want to define a conditional state $f:L\times L_0\to [0,1]$ on $L$, then of course, the fulfillment of \eqref{L-0} is a necessary condition for the existence of $f$. However, it is an open problem if this condition is also sufficient. 
\end{remark}

\begin{definition}\label{def:indep}
Let $L$ be a $\sigma$-OML and let $f(.|.)$  be a conditional state on $L$.
For $a,b\in L$ we say that  \emph{$b$  is independent of $a$ with
respect to the state $f(.|{\mathbf 1}_L)$} %($b\asymp_{f(.|{\mathbf 1}_L)} a$) 
if $f(b|{\mathbf 1}_L)=f(b|a)$.
\end{definition}

In fact, $f(\cdot|{\mathbf 1}_L)$ plays the role of a prior state
(prior probability distribution) in the classical definition of
independence. This means that Definition \ref{def:indep} is just
re-written from the Kolmogorovian probability theory. But unlike
the Kolmogorovian theory, the independence of elements of a $\sigma$-OML
is not necessarily symmetric. In \cite{Nc} a
conditional state $f$ was constructed in such a way that there are elements
$a,b\in L$ for which $f(b|{\mathbf 1}_L)=f(b|a)$ and $f(a|{\mathbf
1}_L)\neq f(a|b)$ (see also Example \ref{exam-smap} later in this paper). This fact implies that the well-known Bayes
Theorem may be violated on a $\sigma$-OML.

\begin{definition}\label{def:smap}
Let $L$  be a $\sigma$-OML. A map $p:L\times L\to [0,1]$  will be called
\emph{an s-map on $L$} if the following conditions are fulfilled:
\begin{enumerate}
 \item[\rm(s1)] $p({\mathbf 1}_L,{\mathbf 1}_L)=1$;
\item[\rm(s2)] for all $a,b\in L$ if $a\perp b$ then $p(a,b)=0$;
 \item[\rm(s3)] for an arbitrary sequence $\{a_i\}_{i\in I}$ of elements of $L$ and arbitrary $b\in L$ if $a_i\perp a_j$ for $i\ne j\in I$, then
\[p\left(\bigvee_{i\in I}a_i, b\right)=\sum\limits_{i\in I}p(a_i,b)\quad\mbox{\rm and}\quad
p\left(b,\bigvee_{i\in I}a_i\right)=\sum\limits_{i\in I}p(b,a_i).
\]
\end{enumerate}
\end{definition}

Let $\cal P$ denote  the system of all s-maps on $L$  (for fixed
$L$), which are $\sigma$-additive in both variables. The
relationship between s-maps $p\in{\cal P}$ and conditional states
is given by the following proposition

\begin{proposition}[\cite{Ns,N-K14}]
Let $L$ be a $\sigma$-OML fulfilling property \eqref{L-0}. \\
{\bf (a)} Assume that there exists a conditional state $f:L\times
L_0\to[0,1]$. Then there exists an s-map $p\in{\cal P}$ such that
for all $a\in L$ and all $b\in L_0$ we have
\[
p(a,b)=f(a|b)\,f(b|{\mathbf 1}_L).
\]
{\bf (b)} Assume that there exists an s-map $p\in{\cal P}$. Then
there exists a conditional state $f:L\times L_0\to[0,1]$ if and
only if for all $b\in L_0$ there exists a $\sigma$-additive state $m_b:L\to[0,1]$
with $m_b(b)=1$. In such a case the conditional state $f$ can
be expressed as
\[
f_p(a|b)=\left\{
\begin{array}{ll}
\frac{p(a,b)}{p(b,b)},\quad&\mbox{if $p(b,b)\ne0$,}\\
m_b(a),\quad&\mbox{if $p(b,b)=0$,}
\end{array}\right.
\]
where $m_b$ is an arbitrary $\sigma$-additive state for which $m_b(b)=1$.
\end{proposition}

Let $L$ be a $\sigma$-OML and $p\in {\cal P}$ be an s-map on $L$.
Denote $\mu_p(a)=p(a,a)$ for all $a\in L$. Then the following
statements hold:
\begin{enumerate}
 \item[\rm(p1)]  $\mu_p:L\to [0,1]$ is a $\sigma$-additive state on $L$.
\item[\rm(p2)] For all $a,b\in L$ we have that $p(a,b)\leq p(a,a)$. %
\item[\rm(p3)] If $a\leftrightarrow b$, then $p(a,b)=\mu_p (a\wedge b )$.
\item[\rm(p4)] For arbitrary $a,b\in L$ the following equivalence holds
\[
f_p(b|{\mathbf 1}_L)=f(b|a)\quad\Leftrightarrow \quad
p(b,a)=p(b,b)p(a,a).
\]
\end{enumerate}

In what follows, for a given s-map $p\in{\cal P}$ we use the
notation  $\mu_p(a)=p(a,a)$ for all $a\in L$.

\begin{proposition}[\cite{AN}]\label{Jauch-P}
Let $L$ be a $\sigma$-OML and $p$ be an s-map on $L$. If $a,b\in L$ are such
that $\mu_p(a)=\mu_p(b)=1$, then $p(a,b)=p(b,a)=1$ and moreover
$p(a,c)=p(c,a)$ for all $c\in L$.
\end{proposition}

\begin{definition}
Let $L$ be a $\sigma$-OML and $p$ be an s-map on $L$. We  say that an s-map $p$ on $L$ is causal if there exist elements $a,b\in L$ such that $p(a,b)\ne p(b,a)$. In this case  elements $a,b$ are said to be $p$-causal.
\end{definition}

We will use the following notation
\begin{eqnarray*}
{\mathcal P}_S &=&\{p\in\mathcal P; p(a,b)=p(b,a)\quad \forall
a,b\in L\},\\
{\mathcal P}_N&=&\mathcal P\setminus\mathcal P_S.
\end{eqnarray*}
\begin{remark}\rm
The property from Proposition \ref{Jauch-P} of an s-map 
\[\mu_p(a)=\mu_p(b)=1\quad \Rightarrow\quad p(a,b)=p(b,a)=1\]
is called the \emph{Jauch Piron property}. This means that causality (the importance of the order,  $p(a,c)$ or $p(c,a)$) can be achieved only if $p(a,a)\ne1$.\\
${\mathcal P}_S$ contains all non-causal s-maps and ${\mathcal P}_N$ all causal s-maps.
\end{remark}

\begin{example}\label{exam-smap}\rm
Let us consider the orthomodular lattices $L_1$ and $L_2$ from Example \ref{exam-OML}. We will show examples of causal s-maps on these lattices.  First we construct an s-map $p_1:L_1^2\to [0,1]$. Because of additivity of $p_1$ in both variables it is enough to present values $p_1(x,y)$ such that $x,y$ are atoms. Besides of these values we give also such values where $x$ or $y$ is equal ${\mathbf1}$ since $p_1(\cdot,{\mathbf1})=p_1({\mathbf1},\cdot)$ is a univariate state.

\begin{table}
\caption{\label{tab-p1}Values of the s-map $p_1$}
%\begin{center}
\begin{tabular}{c|ccccc}
\hline\noalign{\smallskip}
&$a$ &$a'$ &$b$ &$b'$ &${\mathbf1}$\\ \hline
$a$& $0.3$& $0$& $0.2$& $0.1$& $0.3$\\
$a'$& $0$& $0.7$& $0.3$& $0.4$& $0.7$\\
$b$& $0.15$& $0.35$& $0.5$& $0$& $0.5$\\
$b'$& $0.15$& $0.35$& $0$& $0.5$& $0.5$\\
${\mathbf1}$& $0.3$& $0.7$& $0.5$& $0.5$& $1$\\
\noalign{\smallskip}\hline
\end{tabular}
%\end{center}
\end{table}

Now we construct an s-map $p_2:L_2^2\to [0,1]$. Also in this case we present only values $p_2(x,y)$ such that $x,y$ are atoms and values where $x$ or $y$ is equal ${\mathbf1}$.

\begin{table}
\caption{\label{tab-p2}Values of the s-map $p_2$}
%\begin{center}
\begin{tabular}{c|cccccc}
\hline\noalign{\smallskip}
&$a$ &$b$ &$c$ &$d$ &$e$ &${\mathbf1}$\\ \hline
$a$& $0.2$& $0$& $0$& $0.2$& $0$&  $0.2$\\
$b$& $0$& $0.4$& $0$& $0.1$& $0.3$ &$0.4$\\
$c$& $0$& $0$& $0.4$& $0$& $0$& $0.4$\\
$d$& $0.15$& $0.15$& $0$& $0.3$& $0$ &$0.3$\\
$e$& $0.05$& $0.25$& $0$& $0$& $0.3$& $0.3$\\
${\mathbf1}$& $0.2$& $0.4$& $0.4$& $0.3$& $0.3$& $1$\\
\noalign{\smallskip}\hline
\end{tabular}
%\end{center}
\end{table}

As we can see in Tables \ref{tab-p1} and \ref{tab-p2}, s-maps $p_1$ and $p_2$ are non-symmetric, i.e., they are causal. But there is one significant difference between these two s-maps. For $p_1$ we have
\[
p_1(a,b)\ne p_1(a,{\mathbf1})\cdot p_1({\mathbf1},b),\qquad p_1(b,a)=p_1(b,{\mathbf1})\cdot p_1({\mathbf1},a),
\]
i.e., $a$ depends on $b$ but $b$ is independent of $a$ (in the lattice $L_1$).\\
For $p_2$ we have
\[
p_2(a,d)\ne p_2(a,{\mathbf1})\cdot p_2({\mathbf1},d)\quad\mbox{and}\quad p_2(d,a)\ne p_1(d,{\mathbf1})\cdot p_1({\mathbf1},a),
\]
i.e., both $a$ is dependent on $d$ as well as $d$ is dependent on $a$ (in the lattice $L_2$).
\end{example}

We say that an s-map $p:L\times L\to[0,1]$ is \emph{strongly causal} if it is causal and there exists a pair of elements $a,b\in L$ such that $a$ is dependent on $b$ but $b$ is independent of $a$.

The s-map $p_1$ from Example \ref{exam-smap} is strongly causal. The s-map $p_2$ from that example is causal, but not strongly causal.

\medskip
An important notion for our considerations is also that of a conditional expectation.
\begin{definition}
Let $L $ be a $\sigma$-OML, $p\in \mathcal P$ be an s-map,
$x\in{\mathcal O}^1_p$ an observable, whose expected value exists, and $\mathcal B$ be a Boolean
sub-$\sigma$-algebra of $L$. \emph{A version of conditional
expectation of the observable $x$ with respect to $\mathcal B$} is
an observable $z$ (notation $z=E_p(x|\mathcal B)$) such that
$R(z)\subset \mathcal B$ and moreover $E_{p}(z|a)=E_{p}(x|a)$ for
arbitrary $a\in \{u\in\mathcal B;\mu_p(u)\neq 0\}$.
\end{definition}
Since for arbitrary observable $y$ $R(y)$ is a Boolean sub-$\sigma$-algebra of $L$  we will write
simply $E_p(x|y)=E_p(x|R(y))$.

\begin{remark}\rm
In fact, the conditional expectation $z=E_p(x|\mathcal B)$ is a projection of the observable $x$ into the Boolean $\sigma$-algebra $\cal B$. This means, if we have $z= E_p(x|y)$ then we have $z\leftrightarrow y$. This property implies that the conditional expectation $E_p(x|y)$ behaves exactly as we are used to from the conditional expectation of random variables in the Kolmogorovian probability theory.
\end{remark}

\subsection{Sum of non-compatible observables}
\label{sec:4-2}
For compatible observables $x,y$ on a  $\sigma$-OML $L$ due to Theorem \ref{L-S} there exist a $\sigma $-homomorphism $h$ and real functions $f,g$ such that  $x=h\circ f$ and  $y=h\circ g$.  This means that $x+y$ is defined by 
   $x+y=(f+g)\circ h$. If $x,y$ are
non-compatible then we cannot apply this procedure and $x+y$ does
not exist in this sense. 

In \cite{N-K14} a sum of non-compatible observables was defined.

\begin{definition}[\cite{N-K14}]
Let $L$ be a $\sigma$-OML and $p\in\mathcal P$. A map
$\oplus_p:{\mathcal O}^1_p\times {\mathcal O}^1_p\to {\mathcal O}^1_p$ is called
\emph{a summability operator} if it yields the following conditions 
\begin{enumerate}
 \item[\rm (d1)]  $R(\oplus_p(x,y))\subset R(y)$;
\item[\rm (d2)]  $\oplus_p(x,y)=E_p(x|y)+y$.
\end{enumerate}
\end{definition}
The following basic properties of $\oplus_p$ are proven in \cite{N-K14}.
\begin{proposition}[\cite{N-K14}]\label{suma}
Let $L$ be a $\sigma$-OML, $\mathcal B$ be a Boolean
sub-$\sigma$-algebra of $L$,  and $p\in\mathcal P$. Assume
$x,y\in{\mathcal O}^1_p$. Then  the following statements are satisfied
\begin{enumerate}
 \item[\rm(e1)] if $x\leftrightarrow y$ then $\oplus_p(x,y)\leftrightarrow\oplus_p(y,x)$;
\item[\rm(e2)]  $\oplus_p^\mathcal{B}(x,y)=\oplus_p^\mathcal{B}(y,x)$;
\item[\rm(e3)]
$E_p\left(\oplus_p^\mathcal{B}(x,y)\right)=E_p(\oplus_p(x,y))=E_p(x)+E_p(y)$;
\item[\rm(e4)] if $\sigma (x)=\{x_1,x_2,...,x_n\}$ and
$\sigma (y)=\{y_1,y_2,...,y_k\}$ then
\[E_p(x)+E_p(y)=\sum\limits_{i}\sum\limits_{j}(x_i+y_j)p(x(\{x_i\}),y(\{y_j\}).\]
\end{enumerate}
\end{proposition}

\section{A model of Granger causality on horizontal sums of Boolean algebras}
\label{sec:5}
Before turning our attention to the Granger causality, we should say something on random vectors and stochastic processes as a generalization of random vectors.

\subsection{Random vectors versus vectors of observables}
\label{sec:5-1}
We will deal with a measurable space $(\Omega,{\cal S})$ where ${\cal S}$ is a $\sigma$-algebra of measurable events. Denote ${\cal B}$ the $\sigma$-algebra of Borel subsets of ${\mathbb R}$. A random variable $\xi:\Omega\to{\mathbb R}$ is an ${\cal S}$-measurable function, i.e., for every $B\in{\cal B}$  $\xi^{-1}(B)\in{\cal S}$.

Further, let ${\cal B}^2$ and ${\cal S}^2$ denote the direct products ${\cal B}\times{\cal B}$ and ${\cal S}\times{\cal S}$, respectively. By $\sigma({\cal B}^2)$ and $\sigma({\cal S}^2)$ we will denote the least set $\sigma$-algebra containing the corresponding direct products. 

Let $\xi$ and $\eta$ be ${\cal S}$-measurable functions. In the Kolmogorovian probability theory the random vector $(\xi,\eta)$ is modelled as a bivariate function such that for every $B\in\sigma({\cal B}^2)$ $(\xi,\eta)^{-1}(B)\in\sigma({\cal S}^2)$. This model works perfectly if $\xi$ and $\eta$ are measurable simultaneously (e.g., two parameters measured on the same objects). But also in this case we are usually interested in knowing probabilities for $P(\xi^{-1}(A),\eta^{-1}(B))$ where $A,B\in{\cal B}$. This means that instead of constructing $\sigma({\cal B}^2)$ and $\sigma({\cal S}^2)$ it is enough (might be up to some exceptions) to work with the corresponding direct products ${\cal B}^2$ and ${\cal S}^2$. Thus the model becomes slightly different from the Kolmogorovian one, especially when we extend this consideration to stochastic processes.

A different situation occurs if we consider a random vector $(\xi,\eta)$, but $\xi$ and $\eta$ are not simultaneously measurable. Of course one possibility how to model this situation is to stay within the Kolmogorovian model. In this case we know that $P((\xi,\eta)^{-1}\in A\times B)=P((\eta,\xi)^{-1}\in B\times A)$, where $A,B\in{\cal B}$. Instead of random variables $\xi$ and $\eta$ we can use observables $\xi^{-1}$ and $\eta^{-1}$. The fact that observables $\xi^{-1}$ and $\eta^{-1}$ are not simultaneously measurable, can be interpreted as their non-compatibility. We have seen in Example \ref{exam-smap} that unlike the probability measure, s-maps are not necessarily symmetric. This means, if we denote $a=\xi^{-1}(A)$ and $B=\eta^{-1}(B)$, we might get $p(a,b)\ne p(b,a)$. However, to get non-compatibility, we must leave Boolean algebras and switch to more general structures. We will consider two copies of the $\sigma$-algebra ${\cal S}$ denoted by  ${\cal S}_1$ and ${\cal S}_2$. Assume ${\cal S}_1\cap{\cal S}_2=\{\emptyset,\Omega\}$.  $\emptyset$ and $\Omega$ are the bottom and top elements, respectively, of these two $\sigma$-algebras. This means that we can make their horizontal sum in the same way as we have made it with blocks $B_1$ and $B_2$ in Example \ref{exam-OML} when we constructing the OML $L_1$. The corresponding horizontal sum of ${\cal S}_1$ and ${\cal S}_2$ will be denoted by $\tilde{\cal S}$. In such a way for arbitrary $A,B\in{\cal B}$ we have $(\xi^{-1}(A),\eta^{-1}(B))\in\tilde{\cal S}\times\tilde{\cal S}$ and
$(\eta^{-1}(B),\xi^{-1}(A))\in\tilde{\cal S}\times\tilde{\cal S}$. In this situation we have one s-map $p$ modelling the (possibly non-symmetric) distribution of both vectors of observables, $(\xi^{-1},\eta^{-1})$ and $(\eta^{-1},\xi^{-1})$.

\begin{remark}[Interpretation of the non-symmetric distribution]\label{rem}\rm
We design two different experiments. In experiment Nr. 1 we measure first a parameter corresponding to $\xi^{-1}$ and then $\eta^{-1}$. In the second experiment we change the order of $\xi^{-1}$ and $\eta^{-1}$.
We admit that the relative frequencies of $(\xi^{-1},\eta^{-1})\in A\times B$ and that of $(\eta^{-1}, \xi^{-1})\in B\times A$, might be different (order-dependent).
\end{remark}

\subsection{Modelling of Granger causality}
\label{sec:5-2}
Assume that $\{\mathbb X_t\}_{t\in T}$  is a stochastic process. For every time-stamp $t\in T$, $X_t$ is a ${\cal S}$-measurable random variable where ${\cal S}$ is a Boolean $\sigma$-algebra. If we want to model causality (in the sense of non-symmetric dependence), we have to make the same procedure as above (with random vectors) when we have abandoned Boolean algebras and considered horizontal sums of Boolean algebras, instead.

We will consider $card(T)$ copies of the $\sigma$-algebra  ${\cal S}$, i.e., we will have a family $\{{\cal S}_t\}_{t\in T}$ and we make their horizontal sum. By $\hat{\cal S}$ we denote the resulting horizontal sum. For every time-stamp $t\in T$ and every Borel set $A\in{\cal B}$ we will have $X^{-1}_t(A)\in\hat{\cal S}$.  
Then, for $s\ne t$, $X^{-1}_t$ and $X^{-1}_s$ are non-compatible observables. We know already  that there exists a joint distribution of $X^{-1}_t$ and $X^{-1}_s$ (or equivalently, conditional distribution $f(X^{-1}_s|X^{-1}_t)$ which is interesting especially when $s>t$), and by Proposition \ref{suma}, having the conditional distribution $f(X^{-1}_s|X^{-1}_t)$, there exists also their sum. 

 \medskip\noindent
{\bf Granger causality.} Assume that we have two (not necessarily stationary) stochastic processes, $\{\mathbf X_t\}_{t\in T}$ and $\{\mathbf Y_t\}_{t\in T}$, where $T$ is a set of all possible time-stamps. According to Definitions 2 and 5 in \cite{G2}, $\{\mathbf Y_t\}_{t\in T}$ causes $\{\mathbf X_t\}_{t\in T}$ if $F(X_{t+1}|Y_t)\ne F(X_{t+1})$, where $F(\cdot|\cdot)$ is a conditional distribution function and $F(\cdot)$ is an unconditioned distribution function. 

\medskip
To model causality between stochastic processes $\{\mathbf X_t\}_{t\in T}$ and $\{\mathbf Y_t\}_{t\in T}$, we need to have an equivalent of a measurable space such that for every $t,s\in T$  observables $X^{-1}_t$ and $Y^{-1}_s$ are non-compatible. This means that we need two copies of  $\hat{\cal S}$ and make their horizontal sum. We denote this newly constructed lattice by $\hat{\cal S}_2$. In this way we get that $F_{(X^{-1}_t,Y^{-1}_t)}$ and $F_{(Y^{-1}_t,X^{-1}_t)}$ may be different functions.

In experiment we are not able to distinguish the order  $(X^{-1}_t,Y^{-1}_t)$ and $(Y^{-1}_t,X^{-1}_t)$ if we measure $X$ and $Y$ at the same time stamp. This means that, as we have already commented in Remark \ref{rem}, measuring the non-symmetric causality experimentally has to follow exactly what Granger proposed in \cite{G,G2}. 

\section{Conclusion}
\label{sec:6}
In this paper we have shown parallels between the Granger causality \cite{G} and modelling of causality on horizontal sums of Boolean algebras which is based on s-maps and conditional states \cite{Ns,Nc,NP08}. The basic property of Granger's causality is its non-symmetry, i.e., the ability to distinguish between a cause and its effect. Causality based on s-maps and conditional states on orthomodular lattices (and on horizontal sums of Boolean algebras as special orthomodular lattices) bears the same property of non-symmetry. This non-symmetry is suitable for modelling of causality (dependencies) in  stochastic processes (as we have shown in Section \ref{sec:5}) where we are able, in a natural way, to distinguish the cause and its effect. As we have pointed out in Remark \ref{rem} such non-symmetry (order-dependence) may occur also when measuring two different parameters, $\xi$ and $\eta$, by designing two different experiments -- first measuring $\xi$ then $\eta$, or vice versa.

%\begin{acknowledgements}
%\end{acknowledgements}



\end{document}